\documentstyle[preprint,eqsecnum,prc,aps]{revtex}

\begin{document}
\draft

\title{
A dynamical model for correlated two-pion-exchange in the pion-nucleon
interaction }

\author{
C. Sch\"utz, K. Holinde and J. Speth }

\address{Institut f\"{u}r Kernphysik, Forschungszentrum J\"{u}lich
GmbH, D--52425 J\"{u}lich, Germany}

\author{B.C. Pearce}
\address{Dept.\ of Physics and Math.\ Physics,
The University of Adelaide, Adelaide 5005, Australia}

\author{J.W. Durso}
\address{Dept.\ of Physics, Mount Holyoke College, MA 01075}

\maketitle

\begin{abstract}
A microscopic model for the $N\bar N\to\pi\pi$ process is presented in
the meson exchange framework, which in the pseudophysical region
agrees with available quasiempirical information. The scalar
($\sigma$) and vector ($\rho$) piece of correlated two--pion exchange
in the pion--nucleon interaction is then derived via dispersion
integrals over the unitarity cut. Inherent ambiguities in the method
and implications for the description of pion--nucleon scattering data
are discussed.
\end{abstract}

\section{Introduction}
\label{sec:intro}

The interaction between a pion and a nucleon plays a prominent role in
low and medium energy physics since it is an important ingredient in
many other hadronic reactions, {\it e.g.} pion production in
nucleon-nucleon collisions or scattering of a pion by a nucleus.

Recently we have presented a meson exchange model for $\pi
N$scattering
\cite{Schuetz94} which contains conventional direct and exchange pole
diagrams (Fig.~\ref{fig:diags}(a)...\ref{fig:diags}(d)) plus
$\sigma$-- and $\rho$--exchange terms (Fig.~\ref{fig:diags}(e), (f)),
and is unitarized by means of the relativistic Schr\"odinger equation.
The main difference from former
models~\cite{Pearce91,Lee91,Gross93,Goudsmit93,Goudsmit93b} is the
evaluation of the scalar--isoscalar ($\sigma$) and vector--isovector
($\rho$) terms.  While in
Refs.~\cite{Pearce91,Lee91,Gross93,Goudsmit93,Goudsmit93b} these
contributions are treated as single exchanges with sharp masses, in
Ref.~\cite{Schuetz94} they were viewed as arising from a correlated
pair of two pions in the $J$=0 ($\sigma$) and $J$=1 ($\rho$) $t$
channels (see Fig.~\ref{fig:correlate}). Their contribution was
evaluated by using quasiempirical information about the t-channel
$N\bar{N}\rightarrow\pi\pi$ amplitudes of Fig.~\ref{fig:correlate} in
the pseudophysical region, which has been obtained by H\"ohler {\it et
al.}\cite{Hoehler83} from an analytical continuation of both $\pi N$
and $\pi\pi$ data, and performing a suitable dispersion integral over
the unitarity cut.

In order to build in constraints from soft pion theorems, a subtracted
dispersion relation was used in Ref.~\cite{Schuetz94} for the scalar
contribution.  This leads to a specific feature apparently favored by
the $\pi N$ data: namely, the resulting interaction is repulsive in
$S$ waves but attractive in $P$ waves.  The approach used in
Ref.~\cite{Schuetz94} led to a considerably stronger contribution from
$\rho$ exchange than used in former treatments. On the other hand, by
defining effective coupling constants suitable for a sharp
$\rho$--mass parametrization we found a rather small tensor to vector
ratio of coupling strengths in the physical $t$ region, in line with
values used before in the $\pi N$ system \cite{Pearce91}.

As shown in Ref.~\cite{Schuetz94}, a model based on the diagrams of
Figs.~\ref{fig:diags} and \ref{fig:correlate} results in $\pi N$ phase
shifts in the elastic region that agree well with empirical
information, as do the scattering lengths and the $\pi N$
$\Sigma$-term ($\approx 65$MeV).

Although the approach outlined above and described in detail in
Ref.~\cite{Schuetz94} for evaluating correlated 2$\pi$-exchange is
certainly adequate for free $\pi N$ scattering, problems arise when
this $\pi N$ interaction is used in other areas of physics.  For
example, modifications of the interaction in the nuclear medium, which
come into play when a pion is scattered by a nucleus, cannot be taken
into account.  The study of such effects requires an explicit
field-theoretic description.

The aim of the present work is to provide such an explicit model for
the correlated 2$\pi$- and $K\bar{K}$-exchange process of
Fig.~\ref{fig:correlate}.  This requires as input realistic
$\pi\pi\rightarrow\pi\pi$ and $\pi\pi\rightarrow K\bar{K}$ $T$
matrices, which we have generated from a potential model based
similarly on meson exchange and involving coupling between $\pi\pi$
and $K\bar{K}$ channels (see Fig.~\ref{fig:pipi}).  The use of such a
dynamical model for the $\pi\pi$ interaction will facilitate future
investigation of not only possible medium modifications of the pion
and nucleon legs, but also of the interaction itself.

The paper is organized as follows: In the next section, the
microscopic model for the $N\bar N\to 2\pi$ process is described and
compared to the data in the pseudophysical
region. Section~\ref{sec:piN} deals with the resulting pion--nucleon
interaction terms arising from correlated $2\pi$ exchange and their
implications for the description of empirical $\pi N$ data.
Section~\ref{sec:summary} contains a short summary and outlook.

\section{Microscopic model for the $N\bar N\rightarrow\pi\pi$ process}
\label{sec:model}

We will generate the amplitude for the process of
Fig.~\ref{fig:correlate} by solving the scattering equation
\begin{equation}
  T_{N\bar {N}\rightarrow\pi\pi} = V_{N\bar{N}\rightarrow\pi\pi} +
    \sum_{pp=\pi\pi,K\bar{K}} T_{pp\rightarrow\pi\pi} g_{pp}
    V_{N\bar{N}\rightarrow pp} \label{eq:nntopipia}
\end{equation}
Here $V_{N\bar{N}\rightarrow pp}$ is the transition interaction and
$T_{pp\rightarrow\pi\pi}$ the transition amplitudes from $\pi\pi$ and
$K\bar{K}$ to $\pi\pi$; both will be specified below (we use $p$ to
denote a generic pseudoscalar meson, $\pi$, $K$ or $\bar{K}$).  Eq.~
(\ref{eq:nntopipia}) could be considered to be a four-dimensional
Bethe-Salpeter-type equation.  However, we use the Blankenbecler-Sugar
(BbS) technique\cite{Blankenbecler66} to reduce the dimensionality of
the integral to three, which simplifies the calculation while
maintaining unitarity.  More explicitly, we have, in the c.m.\ system
and in the helicity representation,
\begin{eqnarray}
  \lefteqn{
  \langle \vec{q}\, 00| T_{N\bar{N}\rightarrow\pi\pi}(t)
    | \vec{p}\, \lambda_{N} \lambda_{\bar{N}} \rangle
  = \langle \vec{q}\, 00| V_{N\bar{N}\rightarrow\pi\pi}(t)
    | \vec{p}\, \lambda_{N} \lambda_{\bar{N}} \rangle
  }\nonumber\\
  &+& \sum_{pp} \int d^{3}k\,
    \frac{
      \langle \vec{q}\, 00| T_{pp\rightarrow\pi\pi}(t)
        | \vec{k}\,\, 00  \rangle
      \langle \vec{k}\, 00| V_{N\bar{N}\rightarrow pp}(t)
        | \vec{p}\, \lambda_{N} \lambda_{\bar{N}} \rangle}
    {(2\pi)^{3} \omega_{p}(k) (t - 4\omega_{p}^{2}(k))}
  \label{eq:nntopipib}
\end{eqnarray}
with
\begin{equation}
  \omega_{p}(k) = \sqrt{k^{2} + m_{p}^{2}}\;\; ,
\end{equation}
where $m_{p}=m_{\pi},m_{K}$ for $p=\pi,K$ respectively.  Thus, $k$ is
the magnitude of the three-momentum part $\vec{k}$ of the relative
four-momentum of the intermediate two-meson state.  The four-momenta
of the two intermediate mesons $k_{1}$ and $k_{2}$ are related to
$\vec{k}$ by
\begin{eqnarray}
  k_{1} = \left( \sqrt{t}/2, \vec{k} \right) \nonumber\\
  k_{2} = \left( \sqrt{t}/2, -\vec{k} \right)
\end{eqnarray}
The helicity of the nucleon (antinucleon) is denoted by $\lambda_{N}$
($\lambda_{\bar{N}})$.  We perform a partial wave decomposition by
writing
\begin{equation}
  \langle \vec{q}\, 00| V_{N\bar{N}\rightarrow pp}(t)
    | \vec{p}\, \lambda_{N} \lambda_{\bar{N}} \rangle
  = \frac{1}{4\pi} \sum_{J} (2J+1) d^{J}_{\lambda 0}(\cos \theta)
    \langle 00| V^{J}_{N\bar{N}\rightarrow pp}(q,p;t)
      | \lambda_{N} \lambda_{\bar{N}} \rangle
\end{equation}
with a similar expression for $T_{N\bar N\to\pi\pi}$.
Here, $d^{J}_{\lambda 0}$ are the
conventional reduced rotation matrices, $\theta$ is the angle between
$\vec{p}$ and $\vec{q}$, and $\lambda=\lambda_{N}-\lambda_{\bar{N}}$.
Using these expressions, Eq.~(\ref{eq:nntopipib}) becomes
\begin{eqnarray}
  \lefteqn{
  \langle 00| T^{J}_{N\bar{N}\rightarrow\pi\pi}(q,p;t) | \lambda_{N}
    \lambda_{\bar{N}} \rangle
  = \langle 00| V^{J}_{N\bar{N}\rightarrow\pi\pi}(q,p;t)
    | \lambda_{N} \lambda_{\bar{N}} \rangle
  }\nonumber\\
  &+& \sum_{pp} \int_{0}^{\infty} dk\, k^{2}\,
    \frac{
      \langle 00| T^{J}_{pp\rightarrow\pi\pi}(q,k;t) | 00 \rangle
      \langle 00| V^{J}_{N\bar{N}\rightarrow pp}(k,p;t)
        | \lambda_{N} \lambda_{\bar{N}} \rangle}
    {(2\pi)^{3} \omega_{p}(k) (t - 4\omega_{p}^{2}(k))}
  \label{eq:nntopipic}
\end{eqnarray}

The $N\bar N\to 2\pi$ on-shell amplitudes are related to the
Frazer-Fulco helicity amplitudes $f^{J}_{\pm}$\cite{Frazer60} via
\begin{eqnarray}
  f^{J}_{+}(t) &=& \frac{p_{on} m_{N}}{4 (2\pi)^{2}
    (p_{on}q_{on})^{J}} \langle 00|
    T^{J}_{N\bar{N}\rightarrow\pi\pi}(q_{on},p_{on};t) | \case{1}{2}
    \case{1}{2} \rangle \nonumber\\ f^{J}_{-}(t) &=& -\frac{p_{on}
    m_{N}}{2 (2\pi)^{2} \sqrt{t}(p_{on}q_{on})^{J}} \langle 00|
    T^{J}_{N\bar{N}\rightarrow\pi\pi}(q_{on},p_{on};t) | \case{1}{2}
    (-\case{1}{2}) \rangle \nonumber\\
\end{eqnarray}
with
\begin{eqnarray}
  q_{on} &=& \sqrt{\frac{t}{4} - m_{\pi}^{2}} \nonumber\\
  p_{on} &=& \sqrt{\frac{t}{4} - m_{N}^{2}} \;\;\; .
\end{eqnarray}

\subsection{The $N\bar{N}\rightarrow\pi\pi,K\bar{K}$ transition
potentials}
\label{sec:transition}

The ingredients of the dynamical model for the transition interactions
$V_{N\bar{N}\rightarrow\pi\pi}$ and $V_{N\bar{N}\rightarrow K\bar{K}}$
employed in this paper are displayed graphically in
Fig.~\ref{fig:tranpot}.  The potential $V_{N\bar{N}\rightarrow\pi\pi}$
($V_{N\bar{N}\rightarrow K\bar{K}}$) consists of $N$ and $\Delta$
($\Lambda$ and $\Sigma$) exchange terms plus $\rho$--meson pole
diagrams.  Their evaluation is based on the following spin-momentum
dependent parts of the interaction Lagrangians
\begin{mathletters}
\label{eq:lagrangian}
\begin{equation}
  {\cal L}_{BBp} = \frac{f_{BBp}}{m_{p}}
    \bar{\psi}_{B} \gamma^{5}\gamma^{\mu} \psi_{B}
    \partial_{\mu} \phi_{p}
\end{equation}
\begin{equation}
  {\cal L}_{NN\rho} = g_{NN\rho} \bar{\psi}_{N} \gamma^{\mu} \psi_{N}
    \phi_{\rho,\mu}
    + \frac{f_{NN\rho}}{4m_{N}} \bar{\psi}_{N}
    \sigma^{\mu\nu} \psi_{N} \left(
    \partial_{\mu} \phi_{\rho,\nu} - \partial_{\nu} \phi_{\rho,\mu}
    \right)
\end{equation}
\begin{equation}
  {\cal L}_{N\Delta\pi} = \frac{f_{N\Delta\pi}}{m_{\pi}}
    \bar{\psi}_{\Delta}^{\mu} \left(
    g_{\mu\nu} + x_{\Delta} \gamma_{\mu}\gamma_{\nu} \right)
    \psi_{N} \partial_{\mu} \phi_{\pi}\;\; +\;\; H.c.
  \label{eq:deltalagr}
\end{equation}
\begin{equation}
  {\cal L}_{\rho pp} = g_{\rho pp} \phi_{p} \phi_{\rho}^{\mu}
    \partial_{\mu} \phi_{p}
\end{equation}
\end{mathletters}
Here, $\psi_{B}$ are the field operators for spin-1/2 particles ($N$,
$\Lambda$, $\Sigma$), $\psi_{\Delta}$ is the spin-3/2 $\Delta$-isobar
operator, $\phi_{p}$ are the corresponding operators for pseudoscalar
($\pi$, $K$) mesons, while $\phi_{\rho}$ denotes the $\rho$ meson.
Also, $\sigma^{\mu\nu}=\frac{i}{2}[\gamma^{\mu},\gamma^{\nu}]$.
The $N\Delta\pi$ coupling (Eq. (\ref{eq:deltalagr})) includes
off-mass-shell contributions, whose strength is characterized by the
parameter $x_\Delta$.
For the propagators, we have
\begin{mathletters}
\begin{equation}
  S_{B}(p) = \frac{\not\!p + m_{B}}{p^2 - m_{B}^{2}}
\end{equation}
\begin{equation}
  S_{\Delta}^{\mu\nu}(p) =
    \frac{\not\!p + m_{\Delta}}{p^2 - m_{\Delta}^{2}}
    \left[ -g^{\mu\nu} + \frac{1}{3}\gamma^{\mu}\gamma^{\nu}
            + \frac{2}{3m_{\Delta}^{2}} p^{\mu}p^{\nu}
            - \frac{1}{3m_{\Delta}} \left( p^{\mu}\gamma^{\nu}
                 - p^{\nu}\gamma^{\mu} \right) \right]
  \label{eq:deltaprop}
\end{equation}
\begin{equation}
  S_{\rho}^{\mu\nu}(p) = \frac{
    -g^{\mu\nu} + \frac{p^{\mu}p^{\nu}}{m_{\rho}^{2}}}
   {p^{2} - m_{\rho}^{2}}
\end{equation}
\end{mathletters}
In this work, we omit the non-pole contributions to the spin-3/2
propagator (Eq.~(\ref{eq:deltaprop}) since it is known\cite{Read73}
that their effect can be taken into account by the second term of the
interaction Lagrangian (Eq.~(\ref{eq:deltalagr}))

As usual, the resulting vertex functions are modified by
phenomenological form factors $F$ to account for the extended vertex
structure.  For the baryon exchange diagrams in Fig.~\ref{fig:tranpot}
we choose
\begin{equation}
  F_{BBp}(q^{2}) = \left(
    \frac{n_{BBp} \Lambda_{BBp}^{2} - m_{B}^{2}}
         {n_{BBp} \Lambda_{BBp}^{2} - q^{2}} \right)^{n_{BBp}}
  \label{eq:cutoffa}
\end{equation}
where $m_{B}$ ($q$) is the mass (four-momentum) of the exchanged
baryon (in the BbS framework adopted here, $q^{2}=-\vec{q}\,^{2}$).
The cutoff masses $\Lambda_{BBp}$ and powers $n_{BBp}$ will be
adjusted later.  For the $\rho$--pole diagrams we introduce form
factors at the meson-meson-meson vertices as follows
\begin{eqnarray}
  F_{pp\rho}(q) &=& \left(
    \frac{n_{pp\rho} \Lambda_{pp\rho}^{2} + m_{\rho}^{2}}
         {n_{pp\rho} \Lambda_{pp\rho}^{2} + 4\omega_{p}^{2}(q)}
      \right)^{n_{pp\rho}} \nonumber\\
  &=& \left(
    \frac{n_{pp\rho} \bar{\Lambda}_{pp\rho}^{2}
          - m_{p}^{2} + \frac{m_{\rho}^{2}}{4}}
         {n_{pp\rho} \bar{\Lambda}_{pp\rho}^{2} + \vec{q}^{2}}
      \right)^{n_{pp\rho}}
  \label{eq:cutoffb}
\end{eqnarray}
with
\begin{equation}
  \bar{\Lambda}_{pp\rho} = \left(
    \frac{\Lambda_{pp\rho}^{2}}{4} + \frac{m_{p}^{2}}{n_{pp\rho}}
    \right)^{1/2}
\end{equation}
In order to judge the behavior of these form factors it is
$\bar{\Lambda}_{pp\rho}$ which should be compared with $\Lambda_{BBp}$
of Eq.~(\ref{eq:cutoffa}) or the conventional monopole cutoff
parameters.

The evaluation of the relevant transition potentials based on
Eqs.~(\ref{eq:lagrangian})--(\ref{eq:cutoffb}) is involved but
straightforward. The resulting expressions have to be multiplied by
appropriate isospin factors derived from SU(3).
More details can be found in Ref.~\cite{Hippchen91}.
Some slight modifications occur since we now use the BbS
framework.

\subsection{The $\pi\pi\rightarrow\pi\pi,K\bar{K}$ amplitude}
\label{sec:pipi}

The starting point for the evaluation of $T_{\pi\pi\rightarrow\pi\pi}$
and $T_{K\bar{K}\rightarrow\pi\pi}$ are the driving terms shown in
Fig.~\ref{fig:pipi}.  Such a model, involving the coupled channels
$\pi\pi$ and $K\bar{K}$ was constructed by our group some time
ago\cite{Lohse90} based on time-ordered perturbation theory.  Here we
use a model with essentially the same physical input, which
alternatively uses the BbS technique.  This procedure proved to be
advantageous when studying the scalar form factor of the pion, kaon
and nucleon \cite{Pearce92} since it has the correct analytic behavior
in the unphysical region (below the $\pi\pi$ threshold).  The
interaction Lagrangians used are (again without isospin)
\begin{mathletters}
\begin{equation}
  {\cal L}_{\epsilon pp} = \frac{g_{\epsilon pp}}{2m_{p}}
    \phi_{\epsilon} \partial_{\mu} \phi_{p} \partial^{\mu} \phi_{p}
\end{equation}
\begin{equation}
  {\cal L}_{vpp} = g_{vpp} \phi_{p} \phi_{v}^{\mu} \partial_{\mu}
\phi_{p}
\end{equation}
\begin{equation}
  {\cal L}_{f_{2}pp} = g_{f_2pp}{2\over m_p} \phi_T^{\mu\nu}
  \partial_\mu \phi_p \partial_\nu \phi_p
\end{equation}
\end{mathletters}
where $v$ denotes the vector mesons $\omega$, $\rho$, $\phi$ and
$K^{*}$ while $f_{2}$ is the tensor meson.  As before, form factors
are attached to each vertex.  For $t$-- ($s$--) channel exchanges,
form factors of the form given in Eq.~(\ref{eq:cutoffa})
(Eq.~(\ref{eq:cutoffb})) are used.
For the $s$--channel pole diagrams in our interaction model, bare
masses have to be used. These pole contributions then get
renormalized to reproduce the physical resonance parameters by the
iteration in the scattering equation.
Values for bare masses, coupling
constants (with some constraints from SU(3) symmetry) and cutoff
masses have been adjusted to reproduce the empirical $\pi\pi$ phase
shifts and inelasticities. These parameters are given in Tables
\ref{tab:masses} ---  \ref{tab:kakapar}.
The description of the data is as successful as in
Ref.~\cite{Lohse90}.  Fig.~\ref{fig:pipiresults} shows the phases for
the $J$=0,1 partial waves of relevance in this paper, as well as
the $S$--wave inelasticity around 1 GeV .(In $P$-waves, the
inelasticity is rather small in this energy region.)

\subsection{The model in the pseudophysical region}
\label{sec:pseudophys}

In order to evaluate the $N\bar{N}\rightarrow\pi\pi$ amplitudes it
remains to specify the parameters in the
$N\bar{N}\rightarrow\pi\pi,K\bar{K}$ transition potentials.
 Masses and most coupling constants are not
treated as fit parameters but are taken from other sources, using
SU(3) symmetry arguments wherever possible.
The $\rho NN$ coupling $f^{(0)}_{NN\rho}$ is taken to be equal to the
$\rho\pi\pi$ coupling.
The parameter $x_{\Delta}$
(Eq.~(\ref{eq:deltalagr})), the bare tensor/vector coupling constant
ratio $\kappa_{\rho}^{(0)}\equiv f_{NN\rho}^{(0)}/g_{NN\rho}^{(0)}$
and the cutoff masses $\Lambda_{NN\pi}$, $\Lambda_{N\Delta\pi}$ have
been adjusted to the quasiempirical results obtained by H\"ohler {\it
et al.}\cite{Hoehler83} from analytic continuation of $\pi N$ and
$\pi\pi$ data. The values used for the baryon exchange contributions
are given in Table~\ref{tab:tranparamsa}.
The value used for $\kappa_{\rho}^{(0)}$ is 4.136.
Note that the functional form of the form factors has
been chosen such that the dependence on the power $n$ is quite weak
(the factor $n$ multiplying $\Lambda^{2}$ in Eqs.~(\ref{eq:cutoffa})
and (\ref{eq:cutoffb}) ensures an expansion of $F(q^{2})/F(0)$ in
powers of $q^{2}$ is independent of $n$ up to order $q^{2}$).  We take
$n_{NN\pi}$ ($n_{N\Delta\pi}$) to be 1 (2).  Since the influence of
the $K\bar{K}$ intermediate state is small anyhow, $\Lambda_{N\Lambda
K}$ and $\Lambda_{N\Sigma K}$ are arbitrarily put to 2.5~GeV.  This
rather large value implies that the $K\bar{K}$ contribution as
evaluated here is probably an upper limit.  For consistency, the
parameters at the $\rho\pi\pi$ and $\rho K\bar{K}$ vertex are taken to
be the same as in the $\pi\pi\rightarrow\pi\pi,K\bar{K}$ model
described in the last section.

We mention that the baryon--baryon--meson form factor parameters
should not be expected to agree with values employed in the Bonn
potential\cite{Machleidt87} and its extension to the hyperon--nucleon
case\cite{Holzenkamp89}.  The reason is that for the $t$--channel
baryon exchange process considered here, one is in a quite different
kinematic regime.  The fact that we cannot establish a definite
relation for the cutoff parameters in different kinematic domains is
the price we have to pay for our simplified treatment of the vertex
structure, which makes the form factor depend on the momentum of only
one particle.  This is a general problem, which, in our opinion, is
difficult to avoid, since a reliable QCD calculation of the full
momentum dependence of the vertex does not exist.

There is one amplitude, $f^{0}_{+}$, for the scalar ($\sigma$) channel
whereas there are two, $f^{1}_{+}$ and $f^{1}_{-}$, for the vector
($\rho$) channel.  In Fig.~\ref{fig:nnbarresults} we show the results
in the pseudophysical region ($t\geq 4m_{\pi}^{2}$) obtained from our
dynamical model, for both the real and imaginary parts.

Given that we have only four free parameters ($\kappa_{\rho}^{(0)}$,
$x_{\Delta}$, $\Lambda_{NN\pi}$ and $\Lambda_{N\Delta\pi}$), there is
remarkable agreement with the quasiempirical result \cite{Hoehler83}
in all amplitudes.
Some disagreement occurs in the scalar amplitude, especially at higher
$t$.  Fortunately, as we will demonstrate below, these do not severely
affect our final result, the correlated $\pi\pi$ (and $K\bar{K}$)
exchange potential in $\pi N$ scattering.  Furthermore one should keep
in mind that the quasiempirical result is subject to considerable
uncertainty at large values of $t$.

\section{$\pi N$ interaction arising from correlated $2\pi$ exchange}
\label{sec:piN}

In order to derive the effective $\sigma$-- and $\rho$--exchange
potentials we use the same procedure as in Ref.~\cite{Schuetz94};
namely, we first perform dispersion integrals over the unitarity cut
using as input the $N\bar N\to\pi\pi$ amplitudes derived in the
foregoing section. Corresponding $\pi N$ potentials are then obtained
in a straightforward way. We refer the reader to Ref.~\cite{Schuetz94}
for details.

\subsection{The potential in the scalar channel}
\label{sec:scalar}

Here, a subtracted dispersion relation is used to impose the chiral
symmetry constraint at the Cheng--Dashen point, with
$\tilde{f}^0_+(2m_\pi^2)$ put to zero; i.e.
\begin{equation}
  \frac{\tilde{f}^{0}_{+}(t)}{t - 4m_{N}^{2}}
    = \frac{t - 2m_{\pi}^{2}}{\pi} \int_{4m_{\pi}^{2}}^{t_c}
    \frac{\mbox{Im} f^{0}_{+}(t')}
        {(t' - t) (t' - 4m_{N}^{2}) (t' - 2m_{\pi}^{2})} \, dt'
  \label{eq:f0disperse}
\end{equation}
with $t_c=50m_\pi^2$. Due to the slightly different $Im f^0_+$
predicted by the dynamical model compared to the pseudoempirical data
of Ref.~\cite{Hoehler83} (see Fig.~\ref{fig:nnbarresults}) the
resulting potential is now a bit stronger compared to that obtained
in Ref.~\cite{Schuetz94}. This is demonstrated in
Fig.~\ref{fig:sigpot}, for the on--shell case and some selected
partial waves.

\subsection{The potential in the vector channel}
\label{sec:vector}

As in Ref.~\cite{Schuetz94} we first start from
\begin{equation}
  \tilde{f}^{1}_{\pm}(t) = \frac{1}{\pi} \int_{4m_{\pi}^{2}}^{t_c}
    \frac{\mbox{Im} f^{1}_{\pm}(t')}{t' - t} \, dt' \;\;\; .
  \label{eq:f1disperse}
\end{equation}
As expected from the excellent agreement of our model amplitudes
$f^1_\pm$ with the quasiempirical ones of Ref.\cite{Hoehler83}
(cp. again Fig.~\ref{fig:nnbarresults}), the present results for
the $\pi N$ potential in the $\rho$--channel are practically the same
as those obtained in Ref.~\cite{Schuetz94}.

However, it was already pointed out in Ref.~\cite{Schuetz94} that
there is a considerable ambiguity in this result. Alternatively,
we could have used a method proposed by Frazer and Fulco
\cite{Frazer60} and applied by H\"ohler and Pietarinen
\cite{Hoehler75}. Here, one first constructs combinations
$\Gamma_{1,2}(t)$ corresponding to vector ($\Gamma_1$) and tensor
($\Gamma_2$) coupling amplitudes
\begin{mathletters}
\label{Gamma}
\begin{equation}
  \Gamma_{1}(t) = - \frac{m_{N}}{p_{on}^{2}} \left(
    f^{1}_{+}(t) - \frac{t}{4\sqrt{2}m_{N}} f^{1}_{-}(t) \right)
\label{Gamma1}
\end{equation}
\begin{equation}
  \Gamma_{2}(t) = \frac{m_{N}}{p_{on}^{2}} \left(
    f^{1}_{+}(t) - \frac{m_{N}}{\sqrt{2}} f^{1}_{-}(t) \right) ,
\label{Gamma2}
\end{equation}
\end{mathletters}
and then performs the dispersion integrals over the unitarity cut,
\begin{equation}
\tilde{\Gamma}_{1,2}(t) = {1\over\pi}\int_{4m_\pi^2}^{t_c}
{Im \Gamma_{1,2}(t')\over t'-t}dt' \;\;\; .
\label{eq:gammadisp}
\end{equation}
Differences in the resulting potentials originate from the additional
$t$--dependence in $\Gamma_{1,2}$ compared to $f^1_\pm$. Despite this
fact, since $\Gamma_{1,2}$ have the same analytic structure as
$f^1_\pm$, both methods would in principle lead to the same results
provided all cut contributions would be taken into account in the
dispersion integrals. Indeed, diagrams included in correlated
two--pion exchange also give rise to left hand cuts. In the example
shown in Fig.~\ref{fig:rhobox} the $N\rho$ intermediate state is the
source of a branch cut in the complex $t$ plane extending from
$-\infty$ to $\simeq -70.5 m_\pi^2$. In fact, there is an infinite
number of such left hand cuts generated by all processes contributing
to correlated correlated two--pion exchange  and it is by far
impossible to include these pieces. Anyhow, $\rho$--exchange is
{\it defined} by the integral over the unitarity cut only.
Therefore it is unavoidable that the results induced by
Eqs. (\ref{eq:f1disperse}) and (\ref{eq:gammadisp}), respectively, will
differ. (Cutting off the integration over the unitarity cut at
$t_c$ turns out to play a minor role only.)

These differences can be nicely demonstrated by parametrizing the
resulting potentials in terms of effective $t$--dependent
$\rho$--coupling strengths $g_{1,2}(t)$ defined by
\begin{equation}
g_{1,2}(t) = 12\pi (m_\rho^2-t)\Gamma_{1,2}(t) \;\;\; ,
\label{eq:g12}
\end{equation}
where $\Gamma_{1,2}$ is either obtained by inserting $\tilde{f}^1_\pm$
calculated using Eq.~(\ref{eq:f1disperse}) into Eqs.~(\ref{Gamma})
or alternatively by dispersing $\Gamma_{1,2}$
(cf. Eq.~(\ref{eq:gammadisp})).
(For the motivation of the definition of $g_{1,2}$,
see Ref.\cite{Schuetz94}.)
In Fig.~\ref{fig:geff} we have plotted the effective vector coupling
strength $g_1(t)/4\pi$, the effective tensor coupling strength
$g_2(t)/4\pi$ and their ratio $\kappa ={g_2\over g_1}$, choosing
$m_\rho$=770 MeV. Since the $t$--dependence in $p_{on}^2$ of
Eq.~(\ref{Gamma}) is rather weak, the resulting $g_2$ does not differ
much. But the factor of $t$ in $\Gamma_1$ leads to a much smaller
$g_1$ if $\Gamma_{1,2}$ are dispersed.

\subsection{Implications for $\pi N$ scattering}
\label{sec:implications}

Our model for correlated $2\pi$ exchange is supplemented by direct and
exchange pole diagrams involving the nucleon and $\Delta$--isobar, and
is then unitarized by means of a relativistic Schr\"odinger
equation. We refer to Ref. \cite{Schuetz94} for details. It has been
shown in that paper that, based on the quasiempirical input for the
$N\bar N\to\pi\pi$ process, a good description of all $\pi N$ partial
waves is obtained by adjusting open form factor parameters.  In that
paper, $\rho$ exchange as defined by Eqs. (\ref{eq:f1disperse}) has
been used.

We first want to discuss what happens when we now replace the
quasiempirical input for correlated $2\pi$ exchange by our dynamical
model. The slight increase in the $\sigma$--channel potential
(Fig. \ref{fig:sigpot}) leads to comparably weakly modified phase
shifts. This effect can be compensated by a small readjustment of the
cutoff parameter (introduced in addition for the $\sigma$ potential,
see Ref. \cite{Schuetz94}), from 1200 MeV to 1120 MeV. There is almost
no change in the $\rho$ channel provided the same ansatz
is used as in Ref. \cite{Schuetz94}.  Therefore a
quantitative description of $S$ and $P$ waves is obtained with
precisely the same values for parameters in pole and exchange diagrams
as in Ref. \cite{Schuetz94} (solid lines in Fig. \ref{fig:SPwaves}).
Corresponding scattering lengths and volumes are given in Table
\ref{tab:sl}.

However, a dramatic change occurs if the $\rho$--exchange potential is
evaluated using Eq. (\ref{eq:gammadisp}).
There is a strong reduction in the $S_{11}$ phase shift
predictions, with smaller modifications in other partial waves (dashed
lines of Fig. \ref{fig:SPwaves}). The latter can be eliminated by
suitably readjusting parameters in the pole and exchange diagrams, but
the discrepancy in $S_{11}$ essentially remains.

In view of this situation, one may ask if the $\pi N$ data can
discriminate between the different formulations for $\rho$
exchange. Within the strict confines of our model, it could be argued
that it does.  On the other hand, the discrepancy could be an
indicator of the absence of an important ingredient still missing in
the $S_{11}$ interaction.  Indeed, there is empirically
well--established resonant structure in that partial wave at higher
energies, which cannot be reproduced by either model. One source for
the required additional attraction in $S_{11}$ is the strong coupling
of this partial wave to the reaction channel $\eta N$.  A second
source of attraction is provided by $N^*_{S11}$ (1535, 1650) pole
diagrams in the $\pi N$ interaction.  If direct coupling of the form
\begin{equation}
{\cal L}_{N^*N\pi} = g_{N^*N\pi} \bar\Psi_{N^*} \vec\tau \Psi_N \vec
                   \Phi_\pi + H. c.
\label{Lagrangian}
\end{equation}
is assumed at the $N^*N\pi$ vertex this process gives rise to
attraction in the $S_{11}$ partial wave of $\pi N$ scattering starting
from the $\pi N$ threshold.

To demonstrate the power of these additional degrees of freedom, in
Fig.~\ref{fig:refit} the result of a simple calculation starting from
the second model for $\rho $ exchange is plotted where an additional
$N^*$ pole diagram has been included. (The parameters used here are:
$m^0_{N^*}$=1550 MeV, $(g^{(0)}_{N^*N\pi})^2/4\pi$ = 0.1,
$\Lambda_{N^*}$ = 2000 MeV with the form factor parametrization of
Ref.~\cite{Schuetz94}.)  Obviously such a model can describe low
energy $\pi N$ scattering.  Therefore, to discard the second model of
$\rho$ exchange on the basis of the current discrepancies is certainly
not justified.

\section{Summary}
\label{sec:summary}

We have presented a dynamical model for the $N\bar N\to\pi\pi$ process
in the meson exchange framework, which in the pseudophysical region
agrees with available quasiempirical information. The scalar
($\sigma$) and vector ($\rho$) piece of correlated two--pion exchange
in the pion--nucleon interaction is then derived via a dispersion
integral over the unitarity cut. Concerning $\rho$ exchange, there is
a sizeable ambiguity in the prediction for its effective strength,
which is due to different formulations of the coupling to the
nucleon. While the restricted low--energy model we have used favors
one formulation, mechanisms such as coupling to the $\eta N$ channel
and inclusion of higher $N^*$ resonances, not treated in our model but
necessary to explain the data at higher energies, could significantly
alter this result, and suggest a direction of future investigation.

\begin{figure}
\caption{Diagrams included in the $\pi N$ potential.}
\label{fig:diags}
\end{figure}

\begin{figure}
\caption{Correlated $\pi\pi$ ($K\bar K$) exchange contributions.}
\label{fig:correlate}
\end{figure}

\begin{figure}
\caption{The contributions to the potential of the coupled channel
$\pi\pi-K\bar{K}$ model.}
\label{fig:pipi}
\end{figure}

\begin{figure}
\caption{The ingredients of the $N\bar{N}\rightarrow\pi\pi,K\bar{K}$
transition potentials}
\label{fig:tranpot}
\end{figure}

\begin{figure}
\caption{$\pi\pi$ phase shifts obtained for $J=0$ and $J=1$
from our coupled channel $\pi\pi -K\bar K$ model and the
$S$--wave inelasticity.
For references to
the data, see Ref.\protect\cite{Lohse90}.}
\label{fig:pipiresults}
\end{figure}

\begin{figure}
\caption{$N\bar{N}\rightarrow\pi\pi$ helicity amplitudes in the
pseudophysical region. The solid lines denote the imaginary parts of
the model amplitudes and the dashed lines the real parts. Squares and
triangles denote the quasiempirical amplitudes taken from
Ref.\protect\cite{Hoehler83}.}
\label{fig:nnbarresults}
\end{figure}

\begin{figure}
\caption{On--shell potentials in various $\pi N$ partial waves arising
from correlated $2\pi$ exchange in the scalar channel. The solid lines
 are the result if the input from the dynamical model is used, the
 dashed lines are based on the pseudoempirical input given in
\protect\cite{Hoehler79}.}
\label{fig:sigpot}
\end{figure}

\begin{figure}
\caption{A diagram contributing to correlated
two--pion exchange and its cuts.}
\label{fig:rhobox}
\end{figure}

\begin{figure}
\caption{Effective coupling strengths for $\rho$ exchange:
(a) vector coupling strength $g_1 / 4\pi$, (b) tensor coupling
strength $g_2 / 4\pi$, (c) $\kappa = g_2 / g_1$.  The solid lines
denote the results if the dispersion integrals are performed for the
$f$ amplitudes (Eq. (\protect\ref{eq:f1disperse}));
the dashed lines show the results if the form factors
$\Gamma_{1,2}$ are dispersed (Eq. (\protect\ref{eq:gammadisp})).}
\label{fig:geff}
\end{figure}

\begin{figure}
\caption{$\pi N$ scattering phase shifts in $S$ and $P$ waves,
as functions of the pion laboratory momentum. The solid lines
originate from the model using the first ansatz for $\rho$ exchange
(Eq. (\protect\ref{eq:f1disperse})), the dashed lines denote the
results if the second ansatz is used
(Eq. (\protect\ref{eq:gammadisp})). Empirical
information is taken from Ref.~\protect\cite{Koch80}.}
\label{fig:SPwaves}
\end{figure}

\begin{figure}
\caption{$S_{11}$ wave $\pi N$ phase shift, as function
of the pion laboratory momentum. The solid and dashed lines denote the
same models as in Fig.~\protect\ref{fig:SPwaves}. The dash--dotted
line gives the result if an additional $N^*(S_{11})$ pole is
implemented in the model based on Eq.~(\protect\ref{eq:gammadisp}).
Empirical information is taken from Ref.~\protect\cite{Koch80}.}
\label{fig:refit}
\end{figure}

\begin{table}
\caption{Masses used throughout the calculation in MeV.  Bare masses
(denoted by the $(0)$ superscript) appear in the s-channel meson
exchanges.  Isospin-averaged masses are used when appropriate.}
\label{tab:masses}
\begin{tabular}{cdcdcd}
Particle & Mass & Particle & Mass & Particle & Mass \\
\hline
$\pi$ & 139.57 & $\rho^{(0)}$ & 1151.3 & $f_{2}^{(0)}$ & 1710.0 \\ $K$
& 495.82 & $\omega$ & 782.6 & $N$ & 938.926 \\ $\epsilon$ & 1400.0 &
$K^{*}$ & 895.0 & $\Delta$ & 1232.0 \\ $\epsilon^{(0)}$& 1505.0 &
$\phi$ & 1020.0 & $\Lambda$ & 1115.6 \\ $\rho$ & 770.0 & $f_{2}$ &
1270.0 & $\Sigma$ & 1193.0
\end{tabular}
\end{table}

\begin{table}
\caption{Parameters used in the $\pi\pi\to\pi\pi$ potential.}
\label{tab:pipipar}
\begin{tabular}{cdcc}
 Vertex & Coupling Constant & Form factor power & Cutoff \\
 $\alpha\beta\gamma$ & $\frac{g_{\alpha\beta\gamma}^{2}}{4\pi}$ &
 $n_{\alpha\beta\gamma}$ & $\Lambda_{\alpha\beta\gamma}$ (MeV) \\
\hline
 $\pi\pi\rho$, $t$--channel $\rho$ exch. & 2.1 & 2 & 1650 \\
 $\pi\pi\rho$, $s$--channel $\rho$ exch. & 2.1 & 2 & 3300 \\
 $\pi\pi\epsilon$, $s$--channel $\epsilon$ exch. & 0.004 & 2 & 2000 \\
 $\pi\pi f_2$, $s$--channel $f_2$ exch. & 0.040 & 2 & 2000
\end{tabular}
\end{table}

\begin{table}
\caption{Parameters used in the $\pi\pi\to K \bar K$ potential.}
\label{tab:pikapar}
\begin{tabular}{cdcc}
Vertices & Coupling Constant & Form factor power & Cutoff \\
$\alpha\beta\gamma$, $\alpha '\beta ' \gamma$ &
$\frac{g_{\alpha\beta\gamma} g_{\alpha ' \beta ' \gamma}}{4\pi}$ &
$n_{\alpha\beta\gamma}=n_{\alpha '
\beta ' \gamma}$ & $\Lambda_{\alpha\beta\gamma}=\Lambda_{\alpha '
\beta '
\gamma}$ (MeV) \\
\hline
$\pi\bar K K^*$,$\pi K\bar K^*$, $t$--channel $K^*$ exch. & 0.525 & 2
& 1800
\\ $\pi\pi\rho$,$K\bar K\rho$, $s$--channel $\rho$ exch. & 1.050 &
2 & 3300 \\
$\pi\pi\epsilon$,$K\bar K\epsilon$, $s$--channel $\epsilon$ exch. &
0.002 & 2 & 2000\\ $\pi\pi f_2$,$K\bar K f_2$, $s$--channel $f_2$
exch. & 0.020 & 2 & 2000
\end{tabular}
\end{table}

\begin{table}
\caption{Parameters used in the $K\bar K\to K\bar K$ potential.}
\label{tab:kakapar}
\begin{tabular}{cdcc}
Vertex & Coupling Constant & Form factor power & Cutoff \\
$\alpha\beta\gamma$ & $\frac{g_{\alpha\beta\gamma}^{2}}{4\pi}$ &
$n_{\alpha\beta\gamma}$ & $\Lambda_{\alpha\beta\gamma}$ (MeV) \\
\hline
$K\bar K\rho$, $t$--channel $\rho$ exch. & 0.525 & 2 & 3100 \\ $K\bar
K\omega$, $t$--channel $\omega$ exch. & -0.525 & 2 & 3100 \\ $K\bar
K\phi$, $t$--channel $\phi$ exch. & -1.050 & 2 & 3100 \\ $K\bar
K\rho$, $s$--channel $\rho$ exch. & 0.525 & 2 & 3100 \\ $K\bar
K\epsilon$, $s$--channel $\epsilon$ exch. & 0.001 & 2 & 2000 \\ $K\bar
K f_2$, $s$--channel $f_2$ exch. & 0.010 & 2 & 2000
\end{tabular}
\end{table}

\begin{table}
\caption{Parameters used in the $N\bar{N}\rightarrow\pi\pi,K\bar{K}$
transition potentials: $t$ channel baryon exchanges.}
\label{tab:tranparamsa}
\begin{tabular}{cdcc}
 Vertex & Coupling Constant & Form factor power & Cutoff \\
 $\alpha\beta\gamma$ & $\frac{f_{\alpha\beta\gamma}^{2}}{4\pi}$ &
 $n_{\alpha\beta\gamma}$ & $\Lambda_{\alpha\beta\gamma}$ (MeV) \\
\hline
 $NN\pi$ & 0.0790 & 1 & 1780 \\
 $N

\Delta\pi$\tablenote{$x_{\Delta}=-0.847$}  & 0.36 & 2 & 1705 \\
 $N\Lambda K$ & 0.0718 & 1 & 2500 \\
 $N\Sigma K$ &  0.00247 & 1 & 2500
\end{tabular}
\end{table}

\begin{table}
\caption{The scattering lengths and volumes in
units $m_\pi^{-(2L+1)}$.}
\label{tab:sl}
\begin{tabular}{ccc}
 & model & Koch and Pietarinen\protect\cite{Koch80} \\
\hline
$S_{11}$ &  0.165 &  0.173 $\pm$ 0.003 \\
$S_{31}$ & --0.092 & --0.101 $\pm$ 0.004 \\
$P_{11}$ & --0.080 & --0.081 $\pm$ 0.002 \\
$P_{31}$ & --0.042 & --0.045 $\pm$ 0.002 \\
$P_{13}$ & --0.029 & --0.030 $\pm$ 0.002 \\
$P_{33}$ &  0.210 & 0.214 $\pm$ 0.002 \\
\end{tabular}
\end{table}

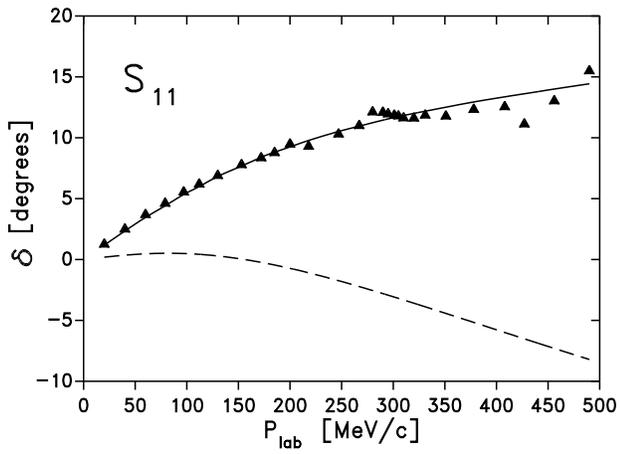
\end{document}